\documentclass[twoside]{article}
\usepackage{epsf,fleqn,espcrc2}
\begin{document}
\title{Benchmarking MILC code with OpenMP and MPI}
\author{Steven Gottlieb and Sonali Tamhankar \thanks{presented by 
S. Tamhankar}\\ 
Indiana University, Bloomington, IN 47405, USA}
\begin{abstract}
A trend in high performance computers that is becoming
increasingly popular is the use of symmetric multiprocessing
(SMP) rather than the older paradigm of MPP.  MPI codes that
ran and scaled well on MPP machines can often be run on
an SMP machine using the vendor's version of MPI.  However,
this approach may not make optimal use of the (expensive) SMP
hardware.  More significantly, there are machines like Blue
Horizon, an IBM SP with 8-way SMP nodes at the San Diego
Supercomputer Center that can only support 4 MPI processes
per node (with the current switch).  On such a machine it is
imperative to be able to use OpenMP parallelism on the node, and
MPI between nodes.  We describe the challenges of converting MILC
MPI code to using a second level of OpenMP parallelism, and benchmarks
on IBM and Sun computers.
\end{abstract}
\maketitle
\section{OpenMP and MPI}
\begin{table*}[t]
\renewcommand{\arraystretch}{1.2} 
\begin{tabular}{|p{1in}|p{2in}|p{2in}|} \hline
 &\hspace{0.8in} OMP &\hspace{0.8in} MPI  \\ \hline
Programming Mode & Additional threads can be created on other processors 
from a master thread. & An MPI process is started on each processor 
that can send and receive messages as needed.  \\ 
Memory & Threads share memory & Each node has separate memory\\ 
Variables & Shared between threads, or private & All variables private 
 \\ 
Global Sum & Builtin Reduction (+) command, which sums a variable 
over all threads & Routines for  global sum\\ 
Communication between CPUs& None needed, since memory is shared & 
MPI send and receive  messages \\ \hline
\end{tabular}
\caption{Comparison between OpenMP and MPI}
\end{table*}
Open MultiProcessing (OpenMP or OMP \cite{omp}) and Message Passing 
Interface (MPI) are two strategies for using multiple processors for 
a single problem. The key difference between them is that in MPI, 
different nodes have their own memory and they communicate with each 
other when needed; but with OpenMP, the memory is shared between threads. 

Here is an example. Suppose we have a two dimensional lattice with 
4 sites in each direction, and we are using four nodes or threads, 
as shown below.
\vspace{0.3cm}\\
\begin{tabular*}{1in}{@{\hspace{0.8in}}cc@{\hspace{0.35in}}cc}
1&2&3&4\\5&6&7&8\\  \\  9&10&11&12\\13&14&15&16\\
\end{tabular*}
\vspace{0.3cm}

Suppose node/thread 1 corresponds to sites 1, 2, 5 and 6. In MPI, 
node 1 has information only about sites on that node, namely 1, 2, 
5 and 6. If it needs information about other sites, for example about 
sites 3 or 7 which are nearest neighbors of sites 2 and 6 respectively, 
it has to use communication routines.

Contrast this with OpenMP, where all threads have access to data for 
all sites, but thread 1 does computations only for sites 1, 2, 5 and 
6.

Table 1 summarizes the differences between the two strategies. A trend 
towards shared memory parallel machines or clusters of Symmetric Multiprocessing 
(SMP) nodes rather than the older paradigm of Massively Parallel Processing 
(MPP) machines makes a study of OpenMP parallelism timely. OpenMP 
was designed to exploit certain characteristics of shared-memory architectures. 
The ability to directly access memory throughout the system (with 
minimum latency and no explicit address mapping) combined with very 
fast shared memory locks, makes shared-memory architectures best suited 
for supporting OpenMP. The advantage of OpenMP is that it is 
easier to program.  Unlike MPI, one does not have to worry about passing 
messages between nodes. In this paper, we study how OpenMP 
performs relative to MPI, and whether combining the two strategies 
gives better performance.

\section{OpenMP DETAILS}
In this section, we give some details about how a C code that works 
for a single processor is changed to work on multiple threads. The 
number of threads is determined by an environment variable, \texttt{OMP\_NUM\_THREADS}. 
The code is executed serially, on a single thread, until a parallel 
construct is encountered, which is executed on multiple threads and 
then serial execution is resumed. To define a parallel construct, 
lines beginning with \texttt{\#pragma omp} are added to the code.
Such pragmas are ignored by the usual C compiler, so the code may also be
run as an ordinary serial code. They are, however, 
interpreted by an OpenMP compiler to identify parallel regions. There 
are several constructs that can be made to execute in parallel; here 
is an example for a ``\texttt{for}'' construct. \vspace{0.3cm}\\ 
\texttt{\#pragma omp parallel for \\   for(i=0;i$<$N;i++)\{\\
my\_job(i);\\\}}\vspace{0.3cm}

This will run the function \texttt{my\_job} in parallel on different 
threads. Note that though the memory is shared, each thread must have 
a private copy of some variables, like \texttt{i} in the above example. 
Loop variables are made private by default but other such variables 
have to be declared ``\texttt{private}''. Some variables may need 
to be summed over all the sites. This is accomplished with a \texttt{reduction} 
statement. The syntax is as follows: \\
\vspace{0.3cm} \\
\texttt{j=0;} \\
\texttt{\#pragma omp parallel for reduction(+:j)}
\\   \texttt{for(i=0;i$<$N;i++)\{\\j+=my\_function(i);\\ \} }\vspace{0.3cm}

This is equivalent to \vspace{0.3cm}\\
\hspace{0.4in}$\texttt{j}=\sum_{\texttt{i=0}}^{\texttt{N-1}} \texttt{my\_function(i)} 
$.\\ \vspace{0.3cm}\\Note that the sum is performed over all threads 
though each thread works only on part of the total number of iterations.

Identifying \texttt{private} and \texttt{reduction} variables is necessary 
for getting correct results.

\section{MILC CODE AND CHANGES}
The MILC \cite{milc} code is a set of publicly available codes developed 
by the MIMD Lattice Computation (MILC) collaboration for doing QCD 
simulations. This code has been run on a variety of
parallel computers, using MPI, for many physics projects.
The files are organized in different directories 
--- the \texttt{libraries} directory contains low level routines like 
matrix multiplication, the \texttt{generic} directory contains oft-needed 
but somewhat higher level routines, including the updating and inversion 
routines. Then there are various application directories. For this 
project, we only concentrated on the conjugate gradient inverter, 
file \texttt{d\_congrad5.c} in \texttt{generic\_ks} directory in version 
6 of MILC code. The code uses a macro \texttt{FORALLSITES} defined 
as \vspace{0.3cm}\\\texttt{\#define FORALLSITES(i,s)$\backslash$ \\ 
for(i=0,s=lattice;i$<$sites\_on\_node;$\backslash$\\i++,s++)}\vspace{0.3cm} 
\\
where \texttt{lattice} is an array of sites, \texttt{site} is a structure 
containing variables defined at each lattice point and \texttt{sites\_on\_node} 
is the number of lattice points on a given node. We needed to redefine 
the macro \texttt{FORALLSITES} because the OpenMP compiler we used 
could not deal with two variables (\texttt{i} and \texttt{s}) in a 
parallel \texttt{for} statement. Here is the macro redefinition.
\vspace{0.3cm}\\ \texttt{\#define FORALLSITES(i,s)$\backslash$\\for(i=0;i$<$sites\_on\_node;i++)\{$\backslash$\\s=\&(lattice[i]);}\vspace{0.3cm}

We used another macro \texttt{END\_LOOP}, which is just defined to 
be a closing brace \texttt{\}} to match the opening brace in the above 
macro.

\section{COMPILING AND RUNNING}
We used the KAP/Pro toolset \cite{kai} for this project. It includes 
the following:
\begin{itemize}
\item{\texttt{guidec}: OpenMP compiler for C.}
\item{\texttt{guideview}: OpenMP parallel performance visualization 
tool. It gives details of program execution, in particular, time spent 
in serial and parallel execution, imbalance in different regions of 
the code, etc.}
\item{\texttt{assurec}: Compiler to be used with debugger which works 
by comparing single thread and multiple thread executions.}
\item{\texttt{assureview}: OpenMP programming correctness tool for 
viewing details of errors or conflicts which occur if different threads 
try to read/write the same variables at the same time.}
\end{itemize}

To add OpenMP parallelism to the MILC code, the following steps were 
required. First, we had to redefine the macro as explained above.  Then we 
added the parallel for pragmas, specifying \texttt{private} 
and \texttt{reduction} 
variables. For example, in the \texttt{FORALLSITES} loop, \texttt{s} 
was made \texttt{private}. We changed \texttt{cc} to \texttt{guidec} 
in our makefiles, and we had to modify those compiler options that 
\texttt{guidec} did not recognize. Adding \texttt{--backend} before 
a compiler option forces \texttt{guidec} to use \texttt{cc} compiler 
options. Then, we ran \texttt{assurec} and \texttt{assureview} to 
locate and remove conflicts. Finally, we ran the executable on different 
number of threads and verified that the output agreed with the MPI 
output.

Even after one has a working OpenMP code, there are some issues to 
consider when comparing its performance with that of MPI.  Some performance 
problems are OpenMP issues, while others are not.  If single thread OpenMP 
performance does not match that with a single node under MPI, that may indicate 
a culprit other than OpenMP.  For example, thread safe compilation 
requires the \texttt{-mt} switch on Sun.  If using this switch on the 
original serial code decreases performance substantially, then the 
performance issue lies with the Sun compiler and its runtime libraries, 
not with OpenMP. If the code uses many \texttt{malloc/free} pairs 
then thread-safe memory allocation is likely the culprit. Again, this 
is not an OpenMP issue, but an issue with the quality of the vendor's 
thread-safe compiler/runtime implementation. We verified that the 
\texttt{-mt} option on the serial version on Sun did not affect the 
performance significantly. Except for the case where we combine OpenMP 
and MPI, no \texttt{malloc/free} statements are used in the region of the code where the performance is
evaluated. Thus, to the best of our knowledge, this is a 
fair comparison between OpenMP and MPI performance. 

\section{RESULTS FOR SUN E10000 AND BLUE HORIZON}
We first ran the modified codes on a Sun E10000 at Indiana University.
The details of architecture for this computer can be
found online \cite{sun}.  Benchmarks were done for various lattice sizes and
numbers of threads.  As the number of threads increased,
the lattice dimensions were increased to keep the volume per thread constant
at $L^4$.  
The number of threads $N$ was increased from 1 to 16 by factors of 2.
For example, for a given $L$, the lattice size for 2 threads is $L^3*2L$ 
and for 16 threads it is $(2L)^4$.  
$L$ was increased from 4 to 14 in steps of 2.
Reported in Fig.~1 is the performance on the Kogut-Susskind quark 
conjugate gradient routine in Megaflop/s per CPU
for both OMP and MPI. For smaller 
number of threads/nodes, the OMP rates are quite comparable to MPI. 
They lag behind for larger number of threads. There are two factors 
involved here: the overhead for setting up threads and the use of the 
cache. For small lattice sizes, since there is only a small number 
of computations to be performed, the former degrades the performance,
but if significant portion of the problem can fit in the cache, the 
execution is speeded up. 
On the other hand, for larger lattices the thread initialization overhead is 
a much smaller fraction of the total computation time, but the problem 
size is too big to fit into the cache. We see that OMP has a ``sweet spot'' 
at size 6, much as the MPI performance peaks at size 8. Since we keep 
the load per thread constant, for the same lattice size the performance 
monotonically decreases in most cases as we increase the number of threads. 
\begin{figure}[thb]
\vspace{-0.2cm}
\epsfxsize=7.0cm
\epsfysize=6.0cm
\epsfbox{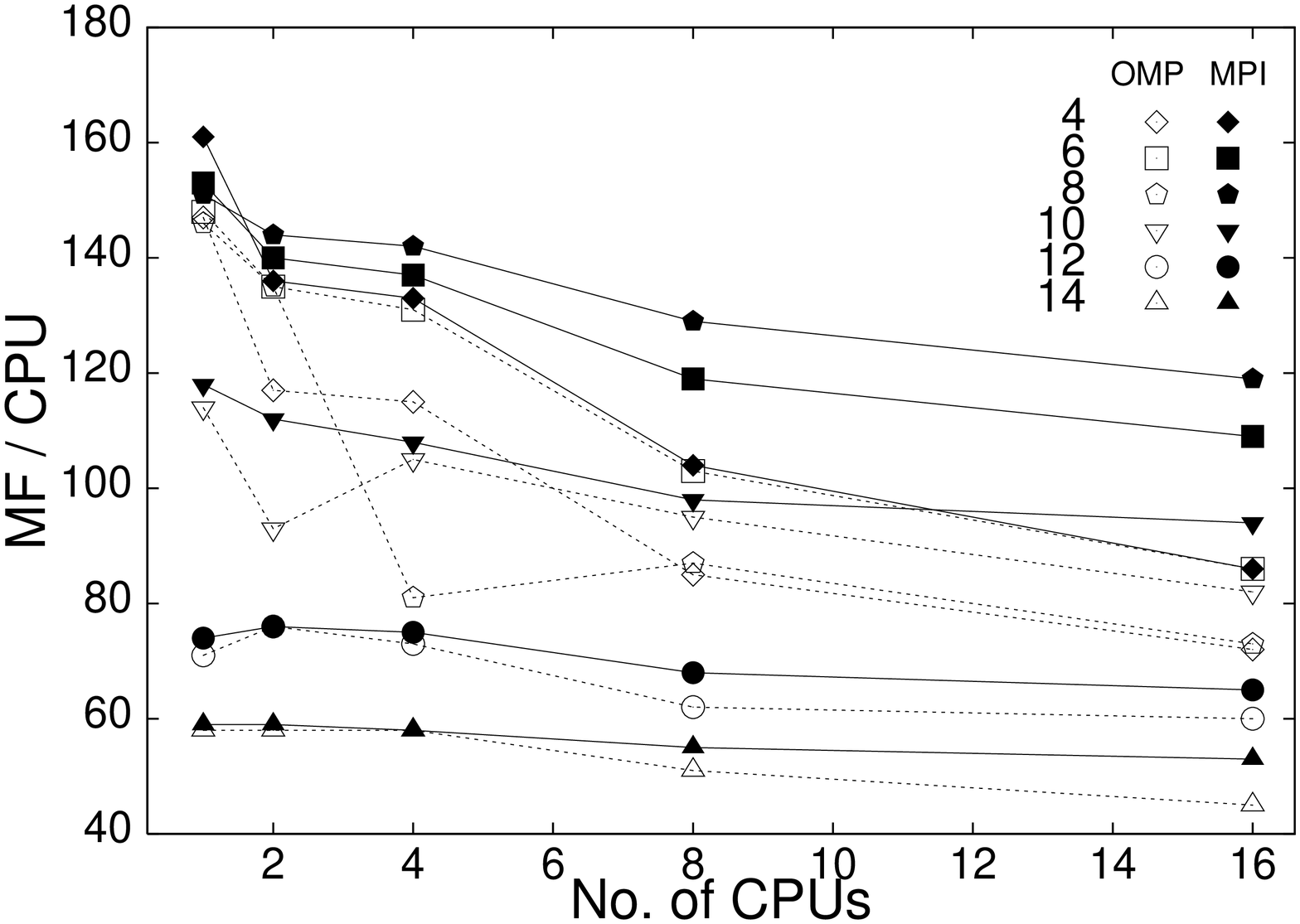}
\vspace{-1.0cm}
\caption{Comparison between OpenMP and MPI performance on Sun E10000. 
The open symbols correspond to OMP and the filled symbols to MPI.}
\label{Solar}
\end{figure}
	
\begin{figure}[thb]
\vspace{0.2cm}
\epsfxsize=7.0cm
\epsfysize=6.0cm
\epsfbox{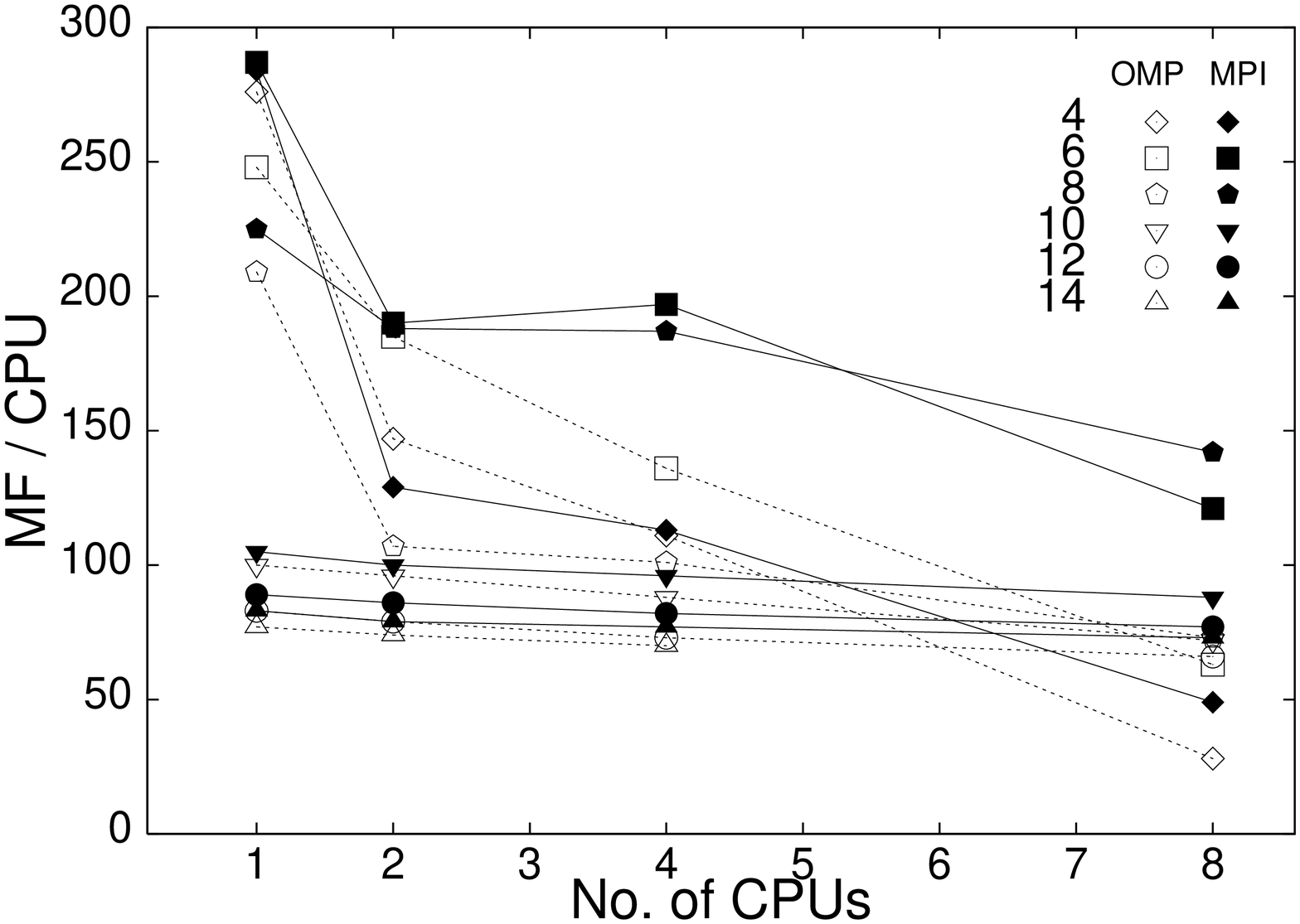}
\vspace{-1.0cm}
\caption{Comparison between OpenMP and MPI performance on Blue Horizon. 
The open symbols correspond to OMP and the filled symbols to MPI.}
\label{Blue}
\end{figure}
Next we benchmarked the code on Blue Horizon \cite{blue}. This IBM 
SP machine at the San Diego Supercomputer Center has 8-way SMP nodes 
but with the current switch can support only 4 MPI processes per node. 
Figure 2 contains the preliminary results from Blue Horizon. These 
results are qualitatively similar to the E10000 results. 

\section{COMBINING OMP AND MPI}
A hybrid approach combining OpenMP parallelism within MPI processes
may offer better performance than either individual approach.
We tried different combinations of threads and MPI processes on Blue 
Horizon.  The hybrid approach fared better at times.  
Figure~3 shows the results for a total of eight processors.
It can be seen 
again that the MPI performance peaks at size 8 (the left-most bars 
in Fig. 3) and the OMP at size 6 (the right-most bars). The combination 
of 2 threads and 4 nodes works best for smaller sizes. The processors 
on Blue Horizon were upgraded after these runs.  We should repeat these
calculations and extend the study to a larger number of CPUs.

\begin{figure}[thb]
\vspace{0.2cm}
\epsfxsize=7.0cm
\epsfysize=6.0cm
\epsfbox{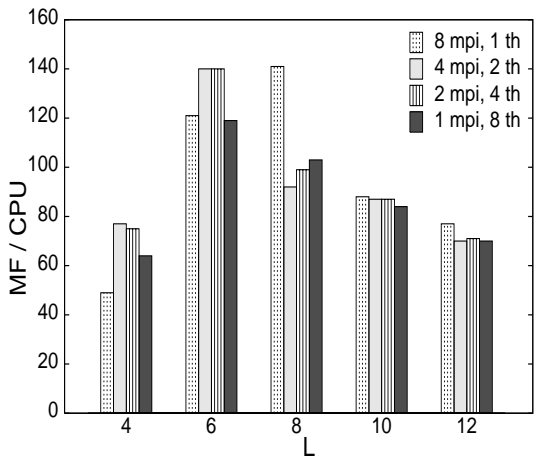}
\vspace{-1.0cm}
\caption{Combining OpenMP and MPI on Blue Horizon.}
\label{combine}
\end{figure}

\section{CONCLUSION}
On both computers studied, OpenMP performance was very similar 
to MPI performance
for a small number of threads, but it deteriorated much faster 
as the number of threads increased, for smaller lattice sizes. Thus, 
OpenMP may be a viable option for someone writing a code to be used with a
modest number of processors on SMP machines. 
The MILC Collaboration, however, already has a working MPI 
code that scales well on many machines. For almost all the combinations 
of problem sizes and number of CPUs studied in this paper, MPI is 
at least as good as OpenMP, if not better.  The only case where we 
get a considerable improvement over MPI is when we combine OpenMP 
and MPI on Blue Horizon for $L=4$ and 6. Not only does the hybrid approach give
the best performance on a single SMP node, it should allow us to run multi-node
jobs using all eight processors on each node rather than the limit of four with
the current switch.
 We have added OpenMP parallelism to the MILC code only for the conjugate 
gradient inverter for this test project. It will require considerably more
effort to modify the whole code to run on multiple OpenMP threads. 
\medskip

It is our pleasure to thank Bill Magro and Henry Gabbs at KAI for 
many useful discussions. 
We gratefully acknowledge the help provided by the staff at Research 
and Technical Services at IU, especially David Hart, Mary Papakhian
and Stephanie Burks. This work was supported by the DOE under grant 
DE-F002-91ER 40661. We thank the San Diego Supercomputer Center and 
NPACI for use of Blue Horizon.

\end{document}